\documentclass[journal]{IEEEtran}

\usepackage{siunitx}
\usepackage[dvips]{graphicx}
\usepackage{bm}
\usepackage{array}
\usepackage{booktabs}
\usepackage{amsmath,amssymb,verbatim,cite,amsopn,graphicx,
url,color,amsthm,enumerate}
\usepackage{algorithm}
\usepackage{algorithmic}
\usepackage{multicol}
\usepackage{epsfig}
\usepackage{epstopdf}
\usepackage{balance}
\usepackage[font={small}]{caption}

\ifCLASSINFOpdf
\else
\fi

\begin{document}
%
\title{Reinforcement Learning based Multi-Access
Control and Battery Prediction with Energy Harvesting in IoT Systems}
%
%
%

\author{
Man Chu,
Hang Li,~\IEEEmembership{Member,~IEEE,}
Xuewen Liao,~\IEEEmembership{Member,~IEEE,}
and Shuguang Cui,~\IEEEmembership{Fellow,~IEEE}
\thanks{This work was accepted in part at the IEEE GLOBECOM, Abu Dhabi, UAE, 2018.
The work was supported in part by grant NSFC-61629101, by NSF with
grants DMS-1622433, AST-1547436, ECCS-1659025, and by Shenzhen Fundamental
Research Fund under Grants No. KQTD2015033114415450 and No. ZDSYS201707251409055.}
\thanks{ M. Chu and X. Liao are with the Department of Information and
Communication Engineering, Xi'an Jiaotong University, Xi'an, 710049,
China  (e-mail: cmcc\_1989414@stu.xjtu.edu.cn, yeplos@mail.xjtu.edu.cn). }
\thanks{H. Li is with the Shenzhen Research Institute of Big Data, Shenzhen, China
(e-mail: hangdavidli@163.com).}
\thanks{S. Cui is with the Shenzhen Research Institute of Big Data and School of
Science and Engineering, the Chinese University of Hong Kong, Shenzhen, China.
He is also affiliated with Department of Electrical and Computer Engineering,
University of California, Davis, CA, 95616. (Email: sgcui@ucdavis.edu).}
\thanks{Corresponding author: S. Cui.}
}
\maketitle

\begin{abstract}
Energy harvesting (EH) is a promising technique to fulfill the long-term
and self-sustainable operations for Internet of things (IoT) systems.
In this paper, we study the joint access control and battery prediction
problems in a small-cell IoT system including multiple EH
user equipments (UEs) and one base station (BS) with limited uplink access
channels.
Each UE has a rechargeable battery with finite capacity.
The system control is modeled as a Markov decision process without complete prior
knowledge assumed at the BS, which also deals with large sizes in both state and
action spaces.
First, to handle the access control problem assuming causal battery and channel
state information,
we propose a scheduling algorithm that maximizes the uplink transmission
sum rate based on reinforcement learning (RL) with
deep Q-network (DQN) enhancement.
Second, for the battery prediction problem,
with a fixed round-robin access control policy adopted,
we develop a RL based algorithm to minimize the prediction loss
(error) without any model knowledge about the energy source and energy arrival
process.
Finally, the joint access control and battery prediction problem is
investigated, where we propose a two-layer RL
network to simultaneously deal with maximizing the sum rate and minimizing
the prediction loss: the first layer is for battery prediction,
the second layer generates the access policy based on the output from the first
layer.
Experiment results show that the three proposed RL algorithms
can achieve better performances compared with existing benchmarks.
\end{abstract}


%
\section{Introduction} \label{section1}

The Internet of things (IoT) has a crucial need
for long-term or
self-sustainable operations to support various
applications\cite{kamalinejad2015wireless}\cite{ni2018dual}.
In recent years, energy harvesting (EH) has been recognized as an emerging
technique that may significantly increase the network lifetime and help reduce
the greenhouse gas emissions in general wireless
applications~\cite{ulukus2015energy}\cite{wang2017beamforming}\cite{chu2017design}.
This technology trend provides a promising energy solution for
IoT applications~\cite{thienself2017}\cite{yang2017energy}.
Accordingly, EH has been being intensively discussed for
supporting the future IoT systems, in D2D communications,
wireless sensor networks, and future cellular
networks~\cite{han2017wirelessly}\cite{tutuncuoglu2012optimum}.
Fundamentally, the amount of harvested energy may be unpredictable due to the
stochastic nature of energy sources,
i.e., energy arrives at random times and in arbitrary amounts,
which poses great challenges to researchers~\cite{chu2017design}\cite{ren2018rf}.
It can be expected that how to handle the dynamics of the harvested
energy would be a key design issue in EH based wireless communication systems.

\subsection{Related Works and Motivations}\label{section1.1}

In general, the related research works on EH based systems could be
categorized into two classes based
on the availability of the knowledge about energy arrivals.
The first class comprises offline approaches that require complete non-causal
knowledge of the considered stochastic system, which are usually adopted to
derive the performance upper
bounds~\cite{ortiz2017multi}\cite{di2017optimal}\cite{yang2012optimal}.
In particular, the optimal uplink resource allocation was
investigated in~\cite{di2017optimal} for the scenario where
two EH users first harvested energy from the wireless signals and then
cooperatively sent information to the access point.
Also, the optimal packet scheduling over multiple access channels
was studied in~\cite{yang2012optimal}, with the
goal of minimizing the time by which all packets from both users are delivered
to the destination.

The second class comprises online
approaches~\cite{blasco2015multi}\cite{khuzani2014online}\cite{blasco2013low}.
Authors in~\cite{blasco2015multi} studied a multi-access wireless system with
EH transmitters, and the access problem was modeled as a partially observable
Markov decision process (POMDP).
In~\cite{khuzani2014online}, the optimal power control policies
for EH nodes in a multi-access system was considered, where a dam
model was constructed to capture the dynamics of the EH process.
In these approaches, some statistic knowledge regarding the dynamic system
should be known at the transmitters~\cite{blasco2013low}.
In many practical applications, the complete non-casual knowledge
or even statistical knowledge of the system dynamics (including both
the channel and energy parts) might not
be available, especially when the EH processes are non-stationary or from sources
with unknown distributions.
For example, in a wireless network with solar EH nodes distributed
randomly over a geographical area, the characteristics of the harvested energy
at each node depend on the node location, and
change over time in a non-stationary fashion~\cite{wang2018combining}.
In such cases, the priori knowledge about dynamics of energy sources is
very difficult to obtain.

Given the above issues, learning based model-free approaches
become more attractive, where the requirements for the priori knowledge are
widely relaxed or even removed~\cite{sakulkar2017online}.
In learning based methods, the learning agent may learn certain statistical
information about an unknown environment system by interacting~\cite{ortiz2017multi}.
In related works, the point-to-point communication with an EH transmitter was
studied in~\cite{blasco2013learning} and~\cite{ortiz2016reinforcement}.
Specifically, a Q-learning based theoretic approach was introduced
in~\cite{blasco2013learning}, where
the transmitter makes a binary decision, i.e., to transmit or not,
in each time slot with the objective of maximizing the total transmitted data.
In~\cite{ortiz2016reinforcement}, the authors studied a transmit power allocation
policy to maximize the throughput using reinforcement learning (RL) with
linear function approximation.
The RL algorithm state-action-reward-state-action (SARSA) was
combined with non-linear function approximation in~\cite{ortiz2016reinforcement}
to enable the use of incoming energy
and channel values, which were taken from a continuous range; thus the authors were
able to improve the performance in an EH point-to-point scenario.
Unfortunately, the theoretical performance cannot be guaranteed and the learning
trends are unstable with non-linear function
approximation, as shown in~\cite{gordon2001reinforcement}.

Given the open nature of wireless systems, there is a crucial need to study
the multiuser systems.
However, most of the existing works
have not provided any stable and efficient learning based approaches for
multiuser access control, especially when the state space and action space
of the considered system are large.
Fortunately, the recently proposed deep Q Network (DQN) technique~\cite{mnih2015human}
successfully adapted the deep neural network as a function approximator in
Q-learning algorithms dealing with large state spaces~\cite{mnih2013playing}.
With DQN, there are two major changes to scale Q-learning: the
network is trained with mini-batch samples from a replay buffer to minimize the
correlations among samples; a target Q network is given to
iteratively update the neural network weights~\cite{gu2016continuous}.

Wireless access control strategies for EH nodes are usually proposed to
make the full use of energy and maintain a perpetual lifetime.
However, the uncertainty of ambient energy availability poses new
challenge to sustainable perpetual operations~\cite{deruyck2018accounting}.
Thus, battery level prediction in such EH based systems is also worth
investigating since a high battery prediction accuracy could potentially
benefit the communication performance.
For example, a novel solar energy prediction
algorithm with Q-learning based on the
weather-conditioned moving average (WCMA) algorithm was proposed in~\cite{kosunalp2016new}.
Unfortunately, this algorithm is restricted to
one type of energy sources and suffers from high computation complexity.
In~\cite{cammarano2016online}, an online energy prediction model
for multi-source EH wireless sensor networks
was proposed, which leverages the past observations to forecast the future
energy availability.
Nevertheless, this energy prediction model requires that the EH dynamics should
be known in advance.
On the other hand, instead of studying the access control and battery prediction
problems separately in EH based IoT applications, it has great
significance to design a joint scheme that feeds the energy prediction results
to the access control design, which could lead to better overall system performances.
This is the focus of this paper.

\subsection{Our Contributions}

To tackle the aforementioned problems,
we focus on an uplink wireless system with $N$ EH user
equipments (UEs) and one BS, where the BS may only use certain causal
information on system dynamics.
We first apply a long short-term memory (LSTM) deep Q-network (DQN) based approach
to design the UE uplink access control.
Then, by fixing the access control policy to be round-robin,
we develop a deep LSTM neural network based battery prediction scheme to
minimize the prediction loss.
Furthermore, we jointly consider the access control and battery prediction
problem using a proposed two-layer LSTM based neural network with
DQN enhancement.
The main contributions are summarized as follows:
\begin{itemize}
 \item We consider an uplink transmission scenario with multiple EH UEs and
 limited access channels, where neither non-casual
 knowledge nor statistical knowledge of the system dynamics (including both
 the channel and energy arrival states) is assumed.
 \item On the condition that only user battery and channel states of
 the current time slot are known at the BS, we propose an LSTM DQN
 based algorithm as the UE uplink access control scheme with the objective of
 maximizing the long-term expected total discounted transmission data.
 Other than the traditional access control problems that
 usually consider maximizing the instantaneous sum rate~\cite{blasco2015multi},
 our goal is to achieve a more stable
 and balanced transmission for a long time horizon.
 \item By fixing the access control policy to be round-robin and
 assuming that the scheduled users embed the information of their true
 battery states in the transmission data, we propose a deep LSTM neural
 network based battery prediction scheme to minimize the prediction loss
 (defined over the differences between the predicted battery states and the true
 battery states within the selected UE set).
 \item We develop a joint access control and battery prediction solution
 by designing 
 a two-layer LSTM DQN network.
 The first LSTM based neural network layer is designed to generate the
 predicted battery levels,
 and the second layer uses such predicted values along with the
 channel information to generate the access control policy.
 The two-layer LSTM based network is trained jointly with the combined objective
 of simultaneously maximizing the total long-term discounted sum rate and
 minimizing the discounted prediction loss of partial users.
 \item The proposed algorithms are designed with many practical considerations
 without strong assumptions.
 In particular, the BS has no prior knowledge on the UEs' energy arrival
 distributions.
 We assume that only the scheduled users embed the information of their
 true battery states in the transmission data, which greatly reduces the
 system signaling overheads.
 \item
 Extensive simulations under
 different scenarios show that the proposed three algorithms
 can achieve much better effectiveness and network
 performance than the various baseline approaches.
\end{itemize}

The rest of this paper is organized as follows.
In Section II, we introduce the system model with some basic assumptions, and also
the
preliminaries on deep Q-learning and LSTM networks.
In Section III, we present the problem formulation for access control, as well
as the LSTM DQN based learning algorithm.
Section IV studies the battery prediction problem.
Furthermore, in Section V, we introduce the joint design problem and its solution.
We provide simulation results in Section VI, and finally conclusions in Section VII.

{\emph{Notations:}}
$s$ and $S_t$ denote state and the state at time slot $t$, respectively;
$a$ and $A_t$ denote action and the action at time slot $t$, respectively;
$R_t$ denotes the reward at time slot $t$;
$\min\{ \text{m, n} \}$ denotes the minimum operator;
$\mathbb{E}_\pi \left[\cdot\right]$ denotes the expected value given that the
agent follows policy $\pi$;
$\log(\cdot)$ denotes the $\log_2(\cdot)$ operator;
$|\cdot|$ denotes the determinant or the cardinality of the set,
depending on the context;
$\|{\cdot}\|$ denotes the $l_2$-norm;
$\nabla$ denotes the first-order derivative operator;
$\mathbb{R}^{M\times{N}}$ denotes the space of real $M\times{N}$ matrixes.

\section{System Model and Preliminaries} \label{section2}

\begin{figure}[!t]
\centering
\includegraphics[width=0.8\columnwidth]{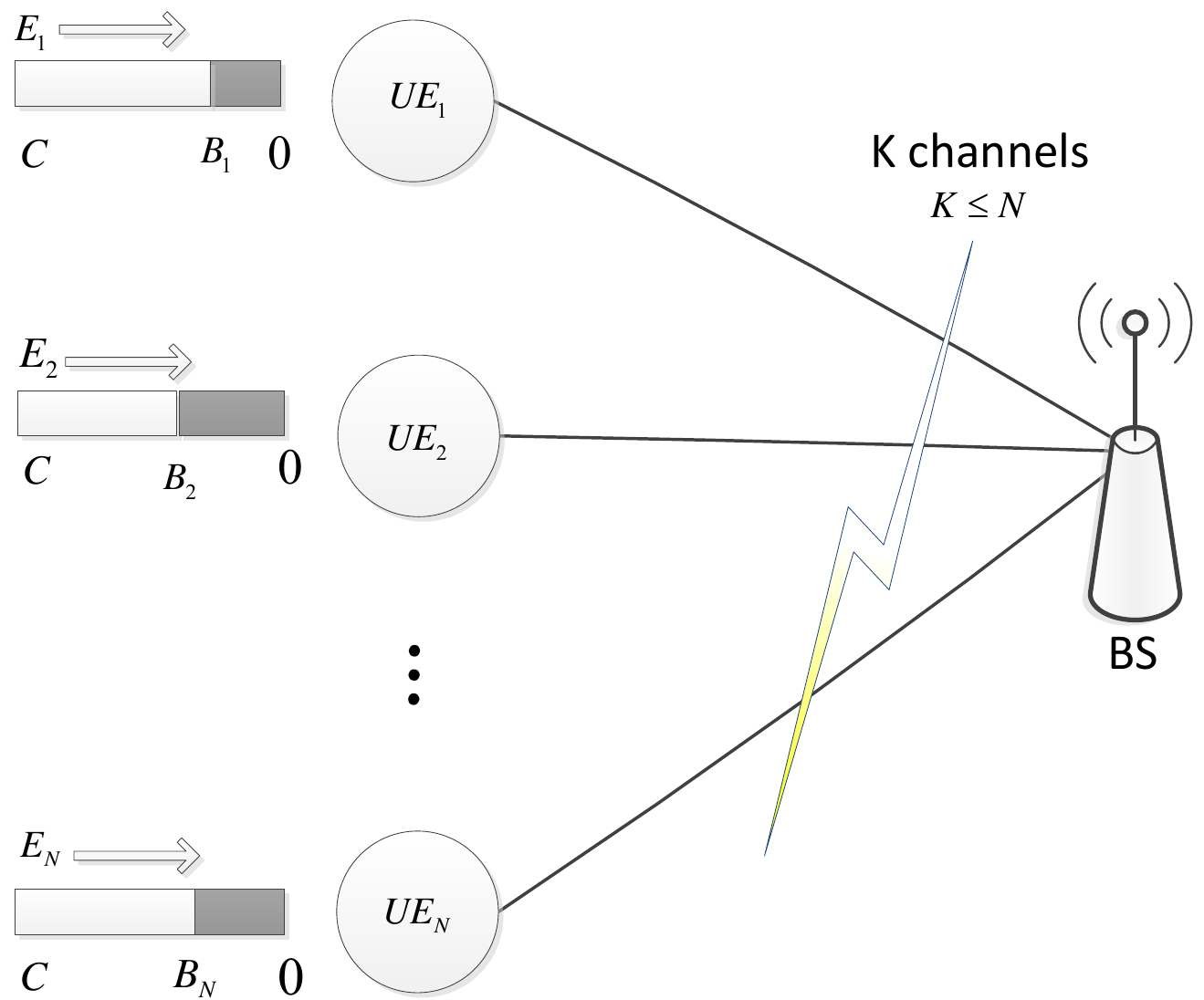}
\caption{System model: $N$ EH users with finite-size batteries
and $K$ orthogonal access channels.}
\label{systemmodel}
\end{figure}

\subsection{System Model}
\label{section21}
We consider an uplink wireless system with $N$ EH based UEs and
one BS, as depicted in Fig.~\ref{systemmodel}.
The system operates in a time-slotted fashion with equal-length time slots (TSs),
with a normalized slot length equal to one.
The BS is able to pick $K$ out of the $N$ UEs to perform uplink access
($K$ can also be viewed
as the number of available orthogonal channels).
The set of all the UEs and the selected subset of UEs at TS $t$ are denoted by
$\mathcal{N}$ and $\mathcal{K}_t$, respectively, where $| \mathcal{K}_t |=K $.
We denote the channel power gain between UE $i$ and the BS at TS $t$ by
${H}_{it}$ and let $\bm{H}_t = \{ {H}_{1t}, \cdots, {H}_{Nt}\}$ 
be the set of all the channel gains at TS $t$.
We assume that at the beginning of each TS, the instantaneous channel state information
(channel power gain) can be obtained at BS.
Besides, the channel states remain constant during each TS and may change across
different TSs~\cite{ortiz2017multi}.
We assume that the UEs always have data for uplink transmission in each TS.
The location of the BS is fixed, while the UEs follow random walks across different TSs
and their locations remain unchanged during one TS.

We assume that all the UEs have no other power sources and
they only use the harvested energy, which is collected
from the surrounding environment via some renewable energy sources (i.e., wind
power, solar power or hydropower).
Many energy arrival processes based on such sources are shown to be
Markov processes~\cite{kosunalp2016new}\cite{cammarano2016online}\cite{yildiz2018hybrid},
but this is not strictly required in our paper.
All the battery states are quantized for analysis convenience~\cite{blasco2015multi}.
We assume that the battery capacity is $C$, same for all the UEs.
We use $E_{it} \in \mathbb{R}$ and $B_{it} \in \{ 0,\cdots, C \}$ to denote the
amount of harvested energy and the state of battery for UE $i$ at the beginning of
TS $t$, respectively.
Let $\bm{B}_t = \{{B}_{1t}, \cdots, {B}_{Nt}\}$ denotes the set of all the
UEs' current battery states.
We assume that the transmission power for each
selected UE is fixed to be $P$~\cite{blasco2015multi}\cite{blasco2013learning}.
After $E_{it}$ is harvested at TS $t$, it is stored in the battery and
is available for transmission in TS $t+1$.
The rechargeable battery is assumed to be ideal, which means that no energy is lost
with energy storing or retrieving and the transmission of data is the
only source of UE energy consumption.
Once the battery is full, the additional harvested energy will be abandoned.
We assume that the power required to activate the UE
EH circuit is negligible compared with the power used for signal
transmission~\cite{chu2017design}\cite{nasir2013relaying}
\cite{zhang2018energy},
since the processing power for EH circuit activation is
usually very small compared
with the transmit power in practice.
For example, as shown in~\cite{zhou2015greedy}, the power
consumption for transmission is about 23 times the
power consumption for activating the EH circuit.

We use a binary indicator $I_{it} \in \{ 0,1 \}$ to describe the access control
policy: If UE $i$ is scheduled to access the channel at TS $t$
(i.e., $i \in \mathcal{K}_t$),
$I_{it} = 1$; otherwise, $I_{it} = 0$.
We use another indicator $z_{it} \in \{ 0,1 \}$ to denote the transmission
status such that: When $P \leq B_{it}$, $z_{it} = 1$, which means that the
transmission could be done successfully;
otherwise, $z_{it} = 0$, which means that at TS $t$, UE $i$
cannot transmit data to the BS and the current transmission is failed.
Based on the above notations, the battery evolution of UE $i \in \mathcal{N}$
over different time slots could be
described as:
\begin{align}
&z_{it}I_{it}P \leq B_{it}\label{tpower}, \\
&B_{i({t+1})} = \min\{C, B_{it} + E_{it} - z_{it}I_{it}P\}.\label{battery}
\end{align}

\subsection{Preliminaries: Deep Q-Learning and LSTM} \label{section23}

In this subsection, we briefly introduce the RL network that is used in this
paper to solve the access control and battery prediction problems.
The detailed physical meanings of the notations in the RL network will be introduced later.
RL is developed over the Markov decision process (MDP) formulation, which includes:
a discrete \textit{state} space $ \mathcal{S} $,
an \textit{action} space $ \mathcal{A} $, an immediate \textit{reward} function
$ r :
\mathcal{S} \times \mathcal{A} \rightarrow \mathbb{R} $ and a transition probability
set $\mathcal{P}$ where $ p(s^{\prime}, r | s, a) \in \mathcal{P}$ satisfies the Markov
property
$ p(s^{\prime}, r|s, a) = \text{Pr}\{S_{t+1}=s^{\prime}, R_t = r|S_t=s, A_t=a\} $,
where $S_t$, $A_t$ and $R_t$ denote the state, action and reward at TS $t$,
respectively.

Note that our paper studies a multiuser uplink scenario.
When we use MDP to model such a system, the number of system states is large
since the state contains the channel gain and battery level for every UE,
and the size of corresponding action space is also large as it is proportional
to the number of UEs.

Based on the MDP formulation, the general goal of an RL agent is to find a good
\textit{policy},
which is a function mapping from state space to the action space, denoted by
$ \pi: \mathcal{S} \rightarrow \mathcal{A}$.
In this paper, the RL agent is the BS, whose goal is to maximize/minimize
the reward/loss in the long run by following the optimal policy.
The total discounted reward from TS $ t $ onwards can be written as:
\begin{equation}
\label{r_t}
R_t^{\gamma} = \sum_{k=t}^{\infty} \gamma^{k-t}R_{k+1},
\end{equation}
where $\gamma \in (0,1)$ is the discount factor.

For a typical RL network,
the \textit{state value function} and \textit{action value function} are instrumental
in solving the MDP, which are defined as
\begin{align}
V_{\pi}(s) = \mathbb{E}_{\pi}[R_t^\gamma \vert S_t = s] =
\mathbb{E}_{\pi} \left[ \sum_{k=t}^{\infty} \gamma^{k-t}R_{k+1} \vert S_t = s \right],
\label{s-value}
\end{align}

\begin{align}
Q_{\pi}(s, a) = &\mathbb{E}_{\pi}[R_t^\gamma \vert S_t = s, A_t = a] \nonumber \\
=&\mathbb{E}_{\pi} \left[ \sum_{k=t}^{\infty} \gamma^{k-t}R_{k+1} \vert S_t = s, A_t = a \right],
\label{q-value}
\end{align}
where $\mathbb{E}_\pi \left[\cdot\right]$ denotes the expected value given that the
agent follows policy $\pi$~\cite{sutton1998reinforcement}.

The optimal policy $\pi^*$ is the policy that can maximize (4) at any state,
and we can observe from (4) and (5) that
$ V_{\pi^*}(s) = \max_{a^\prime \in \mathcal{A}} Q_{\pi^*}(s, a^\prime)$.
The corresponding action-value function for the optimal
policy $\pi^*$ is denoted by $Q_{\pi^*}(s, a)$.
A fundamental property of the value functions is that the functions can be
evaluated in a
recursive manner by using the Bellman equations.
The general form of the Bellman optimality equation for the action value
function is given as
\begin{align}
Q_{\pi^*}(s, a) &= \mathbb{E} \left[ R_{t+1} + \gamma \max_{a^\prime} Q_{\pi^*}(S_{t+1}, a)
\vert S_t = s, A_t = a  \right]  \nonumber \\
&= \sum_{s^\prime}p(s^\prime, r \vert s, a) \left[ r(s,a,s^\prime)
+ \gamma \max_{a^\prime} Q_{\pi^*}(s^\prime, a^\prime) \right],
\label{q-star}
\end{align}
where $r(s,a,s^\prime) = \mathbb{E} \left[ R_{t+1} \vert S_t = s, A_t = a,
S_{t+1} = s^\prime  \right]$ is the expected value of the next reward given the
current state $s$ and action $a$, together with the next
state $s^\prime$~\cite{sutton1998reinforcement}.

In a system with large state and action spaces, it is often impractical
to maintain all the Q-values, i.e., all the function values in~\eqref{q-value}
for all possible state-action pairs.
Generally, nonlinear function approximation
for learning $Q_{\pi}(s, a)$ (Q-learning)
is a popular approach~\cite{ortiz2016reinforcement}.
However, it usually cannot provide theoretical guarantees, producing
unstable trajectories in many practical applications.
Fortunately, the recently proposed DQN~\cite{mnih2015human}
successfully adapts the deep neural network as a function approximator in
Q-learning over large state spaces.
In our work, we adopt the DQN to approximate the action value
function for all state action pairs.

\begin{figure}[!t]
\centering
\includegraphics[width=0.8\columnwidth]{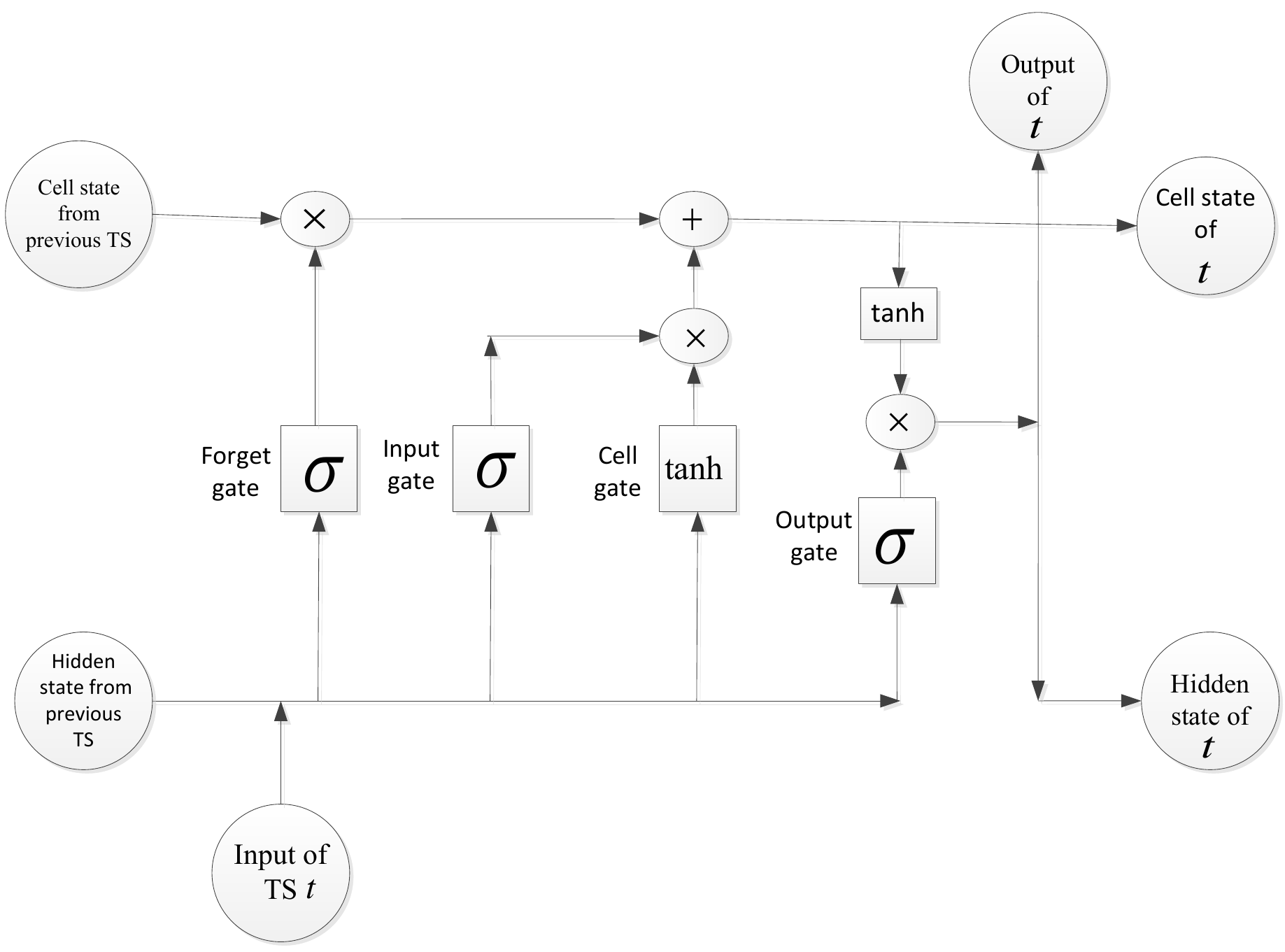}
\caption{LSTM unit.}
\label{lstmblock}
\end{figure}

In particular, we build the DQN based on Long Short-Term Memory (LSTM),
which is a special recurrent neural network that
can connect and recognize long-range correlated patterns over the input and
output states.
Specifically, an LSTM network is considered as multiple copies of the
memory blocks (LSTM units), each of which passes a message to its successor as
shown in Fig.~\ref{lstmblock}.
Such an LSTM unit has four gates to control the flow of information.
With the input from the current step $t$ and the hidden state of the
previous step $t-1$,
the unit firstly decides what information to
throw away through multiplying
the forget gate output by the cell state from $t-1$.
The cell state runs straight down all the units in the
LSTM network.
The next procedure is to decide what new information is going to be stored in the cell
state using the input gate and cell gate.
Finally, the hidden state is updated with
the new cell state and the output of the output gate.

\subsection{Performance Metric}
In this paper, we adopt two main metrics to measure the overall
performance of the network.
The first is the sum rate of all the uplink transmissions to the BS.
At time slot $t$,
the network uplink sum rate at the BS is given by
\begin{align}
\mathcal{R}_t = {\sum_{i \in \mathcal{K}_t}  z_{it} F
\log \left(1 + \frac{P {H}_{it}}{\sigma^2} \right)},
\end{align}
where $F$ is the spectrum bandwidth and $\sigma^2$ is
the noise power~\cite{verdu1998fifty}.

The second metric is the prediction loss, which is the dissimilarity between
the predicted battery states and the true battery states.
It is worth mentioning that in our battery prediction design,
the BS does not know the full information of $\bm B_t$ (i.e., the
battery states at each TS) for decision making.
In order to avoid large signaling overheads for reporting
battery states to the BS at each TS, only the selected UEs send their true
battery states along with transmitted data to the BS.
Thus, the instantaneous prediction loss only involves the selected UE set.
However, as the RL algorithm explores all the UEs, all the UEs will be taken
into account in the long run.
The instantaneous prediction loss $P_{loss}(t)$
at time slot $t$ is given by
\begin{align}
P_{loss}(t) = \sqrt{ \sum_{i \in \mathcal{K}_t}{\lvert B_{it} - b_{it} \rvert}^2 },
\label{ploss}
\end{align}
where $b_{it}$ and $B_{it}$ are the predicted battery state and true battery
state of UE $i$ at time slot $t$, respectively.

\section{Access Control with RL} \label{section3}

In this section, we consider the access control problem
described in Fig.~\ref{systemmodel}.
It is assumed that the BS and UEs are cooperative such that the BS may obtain
the knowledge of current channel gains and UEs' current battery
states~\cite{ortiz2017multi}.
The system operates as follows. When the system enters a new time slot,
the BS uses the current UE battery states and channel information to compute
its scheduling policy for the current TS with a RL network, and then broadcasts
the policy to all the UEs.
Afterwards, the selected UEs transmit their data to the BS with transmission
power $P$, while those who are not selected remain idle.
All the UEs execute the energy conversion process and store the energy into
the battery for the future use.
The above process repeats in the next TS.

\subsection{Problem Formulation}
The BS needs to find a good access control policy with the
objective of maximizing the long-term expected uplink sum rate.
In TS $t$, the system state $S_t$ contains two parts: the current channel state
information $\bm{H}_t = \{{H}_{1t}, \cdots, {H}_{Nt}\}$ and the current
UE battery state $\bm{B}_t = \{{B}_{1t}, \cdots, {B}_{Nt}\}$.
That is, we have $S_t = \{ \bm{H}_t, \bm{B}_t\}$.
The action space $\mathcal{A}$ contains all the possible UE
selection choices, i.e., $\mathcal{K}_t \in \mathcal{A}$,
with $|\mathcal{K}_t| = K$, $\sum_{i=1}^N I_{it} = K$.
Here, the reward signal $R_t$ is the received sum rate at the BS, which is
regarded as the reward at state $S_t$ by taking action $A_t$.
The description of $R_t$ is shown in (7) and the
total discounted sum rate (reward) can be calculated as
\begin{align}
R_t^{\gamma} &= \sum_{k=t}^{\infty}{\gamma^{k-t}}R_{k+1} \nonumber \\
&= \sum_{k=t}^{\infty}\gamma^{k-t}{\sum_{i \in \mathcal{K}_{k+1}}  z_{i(k+1)} F
\log \left(1 + \frac{P {H}_{i(k+1)}}{\sigma^2} \right)}.
\end{align}

Our learning goal is to maximize the expected cumulative discounted reward
by following the access policy $\pi$ from a starting state,
which is given by
\begin{align}
J_1(\pi) = \mathbb{E}(R_t^{\gamma} \mid \pi).
\label{ob}
\end{align}
As such, the access control optimization problem can be formulated as
\begin{subequations}
\label{P1}
\begin{align}
\max\limits_{\pi}\;\;\;
&J_1(\pi)\\
{\rm{s}}{\rm{.t}}{\rm{.}}\;\; & (1)\; \text{and} \;(2) \\
& \sum_{i=1}^N I_{it} = K .
\end{align}
\end{subequations}

\subsection{LSTM-DQN based Access Control Network}

\begin{figure}[!t]
\centering
\includegraphics[width=0.8\columnwidth]{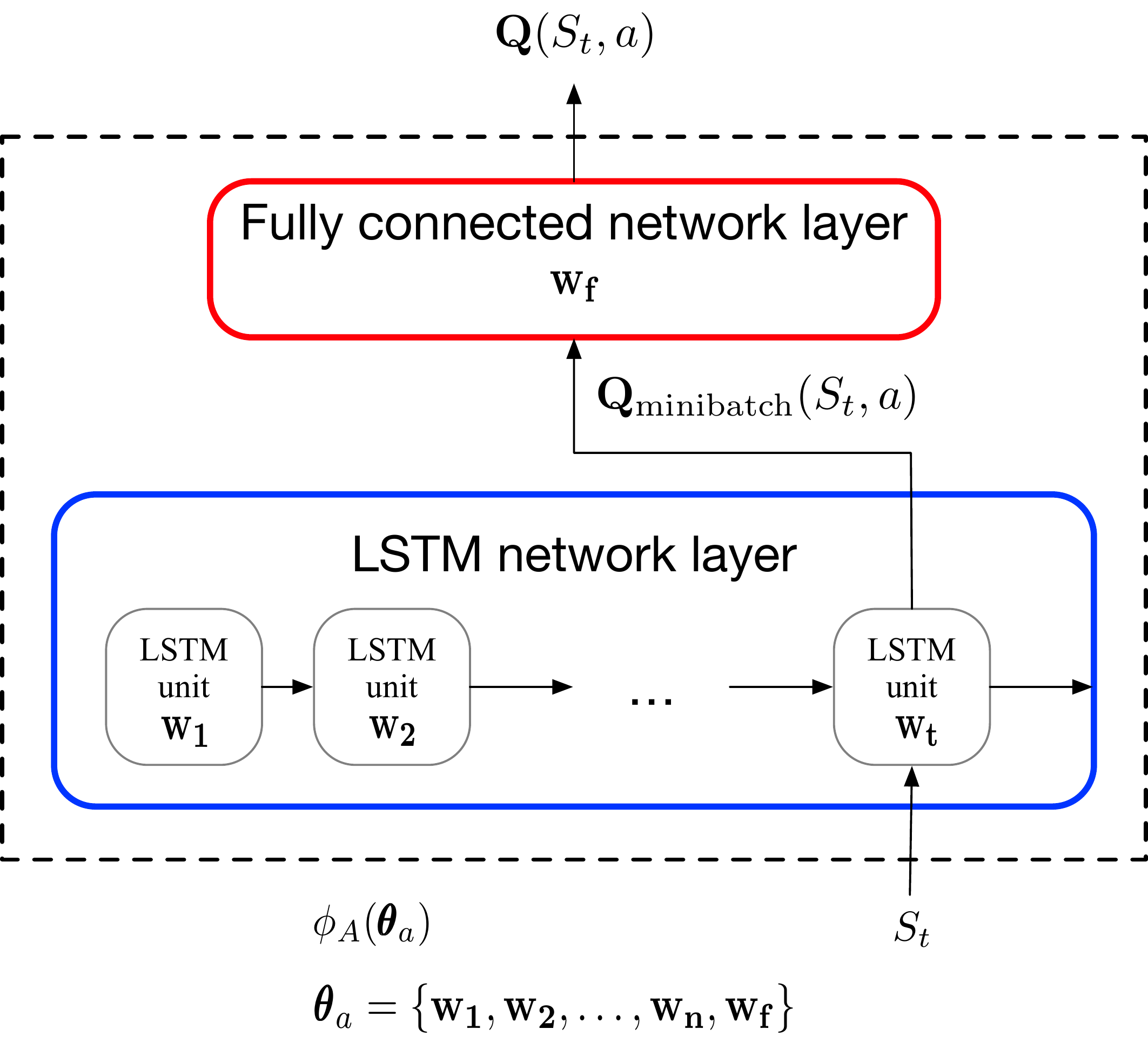}
\caption{LSTM based access control network $\phi_A(\bm \theta_a)$.}
\label{onelstm}
\end{figure}

In this subsection, we present the proposed learning framework and algorithm
of uplink access control to solve problem~\eqref{P1}.
In Fig.~\ref{onelstm}, we illustrate the architecture of our LSTM-DQN network
for access control.
The centralized controller at the BS receives the state information $s$
at the beginning of each TS.
With the input $s$, the entire layer of the LSTM network outputs
the approximated Q-value in mini-batch
$\bm Q_\text{minibatch}(s,a) \in  \mathbb{R}^{\mathrm{batchsize} \times {L}}$,
where $L$ is the size of the action space and $a \in  \mathcal{A}$.
Then, we use a fully connected network layer to adjust the size of the Q-value
vector to make it fit the action space, i.e.,
$\bm Q(s, a) \in \mathbb{R}^{1 \times L}$.
We use $\phi_A(\bm \theta_a)$ to represent our neural network in Fig.~\ref{onelstm}
with the input as the system state $s$ and the output as $\bm Q(s, a)$.
Here, $\bm \theta_a$ denotes the set of network weights which
contains:
the LSTM layer parameters $\{\bm w_1, \cdots, \bm w_n\}$ and the fully
connected network
layer parameters $\bm w_f$, where $n$ is the number of LSTM units.

In the learning time slot $t$, $\bm Q(S_t, a)$ is estimated by $\phi_A(\bm \theta_a)$.
We recall that given $S_t$, with $\bm Q(S_t, a)$ at hand the BS selects $A_t$
that achieves the maximum $Q(S_t, A_t)$, and the optimal policy is the greedy policy
if $\bm Q(S_t, a)$ can be perfectly estimated.
However, the greedy policy is not optimal before the estimate of
$\bm Q(s, a)$ is accurate enough.
In order to improve such estimates, the BS should balance
the exploration of new actions and the exploitation of the known actions.
In exploitation, the BS follows the greedy policy; in exploration the BS takes
actions randomly with the aim of discovering better policies.
The balance could be realized by the $\epsilon$-greedy action selection
method~\cite{narasimhan2015language}
(as given later in Algorithm 1) at each time slot, which either takes actions
randomly to explore
with probability $\epsilon$ or follows the greedy policy to exploit with
probability $1-\epsilon$, where $0 < \epsilon <1$.

After executing the selected action $A_t$, the BS receives the reward $R_t$ and the
system changes to the new state.
We utilize experience replay to store the BS's experiences at each TS, which is
denoted by tuple $e_t = (S_t, A_t, R_t, S_{t+1}) $ in a data-set $\mathcal{D} = \{e_1,
..., e_t\}$.
The replay memory size is set to be $L$, which means their we could store
$L$ experience tuples.
Here, $e_t$ is generated by the control policy $ \pi (a|s) $.
In each TS, instead of updating $\bm \theta_a$ based on
transitions from the current state, we randomly
sample a tuple $ (\tilde s, \tilde a, \tilde r, \hat s) $ from $\mathcal{D}$.
Updating network parameters in this way to avoid issues caused by strong
correlations
among transitions of the same episode~\cite{mnih2015human}.
We parameterize an approximate value function $Q(s, a ; \bm \theta_{a})$ using
the proposed learning network in Fig.~\ref{onelstm} with network
parameters (weights) of $\bm \theta_{a}$.
With the sampled transitions,
$y_t = \tilde r +  \gamma  \max_{\hat a} Q(\hat s, \hat a; \bm \theta_a^-)$
is the target Q-value with network weights $\bm \theta_a^{-}$ obtained from
previous iteration.
Accordingly, we have the following loss function to minimize:
\begin{align}
\mathcal{L}_t(\bm \theta_a) = (y_t - Q(\tilde s, \tilde a ; \bm \theta_a))^2.
\end{align}
Differentiating the loss function with respect to the weights, we arrive the following
gradient:
\begin{align}
\label{eq}
&{\nabla}_{\bm \theta_a} {\mathcal{L}_t(\bm \theta_a) }
=  \nonumber \\
&\left(\tilde r + \gamma \max_{\hat a} Q(\hat s, \hat a; \bm \theta_a^-)-Q(\tilde s, \tilde a ; \bm \theta_a)\right)
{\nabla}_{\bm \theta_a}Q(\tilde s, \tilde a ; \bm \theta_a),
\end{align}
where ${\nabla}_{\bm\theta_a} f(\cdot)$ denotes
the gradient vector of $f$ with respect to $\bm\theta_a$.
By adopting the routine of stochastic gradient descent~\cite{mnih2015human},
the overall access control algorithm is summarized in Algorithm 1.

\begin{algorithm}[h]
\caption{LSTM-DQN based Access Control Algorithm}
  \label{table:1}
\begin{algorithmic}[1]
	\STATE {Initialize the experience memory $\mathcal{D}$, }
	\STATE {Initialize the parameters of action generator network $\phi_A$ with
          random weights $\bm \theta_a$,}
	\STATE {Initialize the total number of episodes $E_p$,}
          \STATE {Initialize the environment and get initial observation $S_1$, }
          \FOR{$t=1, \cdots, \infty$}
             \IF{$random () \leq \epsilon$}
               \STATE {Select a random action $A_t \in \mathcal{A}$;}
             \ELSE
               \STATE {Compute $Q(S_t, a)$ for all actions $a \in \mathcal{A}$ using $\phi_A$,}
               \STATE {Select $A_t = \underset{{a \in \mathcal{A}}}{\arg\max}Q(S_t, a)$.}
             \ENDIF
             \STATE {Execute $ A_t $, observe reward $ R_t $ and new state $ S_{t+1} $,}
             \STATE {Store transition $(S_t, A_t, R_t, S_{t+1})$ in $\mathcal{D}$,}
             \STATE {Sample random mini-batch of transitions $(\tilde s, \tilde a, \tilde r, \hat s)$ from $\mathcal{D}$,}
             \STATE {Set $y_t = \tilde r$ if $t+1$ is the
             terminal step of the episode ($t+1 = E_p$); otherwise,
             $y_t = \tilde r + \gamma \max_{\hat a} Q(\hat s, \hat a; \bm \theta_a^-)$, }
             \STATE {Perform stochastic gradient descent step on the loss function
             $\mathcal{L}_t(\bm \theta_a) = (y_t - Q(\tilde s, \tilde a ; \bm \theta_a))^2$ to
             update network parameters $\bm \theta_a$ according to~\eqref{eq}. }
           \ENDFOR
\end{algorithmic}
\end{algorithm}

\section{Battery Prediction with RL}

In wireless EH network, it is important to keep energy-neutral, i.e., energy
expenditure equals the harvested amount and operation permanently.
An effective energy-neutral policy may benefit from an accurate prediction for
the future harvested energy.
In this section, we consider reusing a similar LSTM method to the one used in the
previous section for UE battery state prediction.
In our design, we do not need to know the UE energy model, as the method
is purely data driven.

\subsection{Problem Formulation}
For the EH system in Fig.~\ref{systemmodel},
we assume that the access control policy is fixed to be the widely used round-robin
scheduling policy~\cite{blasco2015multi}.
At the beginning of each TS $t$, the LSTM based predictor can output the predicted
battery states $b_{it}, i \in \mathcal{N}$.
The BS then schedules the UE transmission based on round-robin scheduling and
broadcasts the schedule to all the UEs.
After receiving the schedule, the selected UEs transmit
data to the BS, along with their current ground-truth battery
states $B_{it}, i \in \mathcal{K}_t$~\cite{blasco2015multi}.
We use the difference between the predicted battery states and the true battery
states within the selected UE set
as the performance metric for the designed predictor, as shown in~\eqref{ploss}.

A memory component with a history window of $W$ is equipped at the BS to store
limited history information.
At TS $t$, the \textit{state} space contains three parts: the
access scheduling history information $\bm{X}_t = \{ I_{ij}\} \in \mathbb{R}^{N \times {W}}$,
where $i \in \mathcal{N}$ and $j \in [t-W+1, t]$; the history
of predicted UE battery information $\bm{M}_t = \{ b_{ij}\} \in \mathbb{R}^{N \times {W}}$,
where $i \in \mathcal{N}$ and $j \in [t-W+1, t]$; and the history of selected UE
true battery information
$\bm{G}_t = \{ B_{ij} \}\in \mathbb{R}^{K \times {W}}$, where
$i \in \mathcal{K}_j$ and $j \in [t-W+1, t]$.

%
%

We use $S_t = \{ \bm{X}_t$, $\bm{M}_t,\bm{G}_t \}$ to denote the system
state, and $\bm{b}_t = [ b_{1t}, \cdots, b_{Nt} ]$ to denote the battery
prediction result, which is also called the prediction
value $\hat {\bm{v}}(S_t) = \bm{b}_t$.
We denote the predicted battery
states of selected UEs as vector $\hat {\bm{v}}_s(S_t)$ with
elements $\{ b_{it} \}_{i \in \mathcal{K}_t}$
and the received true battery states of selected UEs
as vector $\bm R_t $ with elements $\{ B_{it} \}_{i \in \mathcal{K}_t}$.
Thus, the prediction loss in~\eqref{ploss} is equivalent to
\begin{align}
P_{loss}(t) = \sqrt{|| \bm R_t - \hat {\bm{v}}_s(S_t)||^2 }.
\end{align}
The long-term prediction performance, i.e., the total discounted prediction
loss, is given by
\begin{align}
\label{ploss_l}
P_{loss}^{\gamma}(t) = \sum_{k=t}^{\infty}\gamma^{k-t}
\sqrt{|| \bm R_{k+1} - \hat {\bm{v}}_s(S_{k+1})||^2 }.
\end{align}

The goal of the learning algorithm is to obtain the optimal prediction
policy $\pi$ to minimize the cumulative discounted loss,
which is given by
\begin{align}
J_2(\pi) = \mathbb{E}(P_{loss}^{\gamma}(t) \mid \pi).
\end{align}
The battery prediction problem can be formulated as
\begin{subequations}
\label{prediction}
\begin{align}
\min\limits_{\pi}\;\;\;
& J_2(\pi)\\
{\rm{s}}{\rm{.t}}{\rm{.}}\;\; & (1)\; \text{and} \;(2).
\end{align}
\end{subequations}

\subsection{ Deep LSTM based Battery Prediction Network}

\begin{figure}[!t]
\centering
\includegraphics[width=0.8\columnwidth]{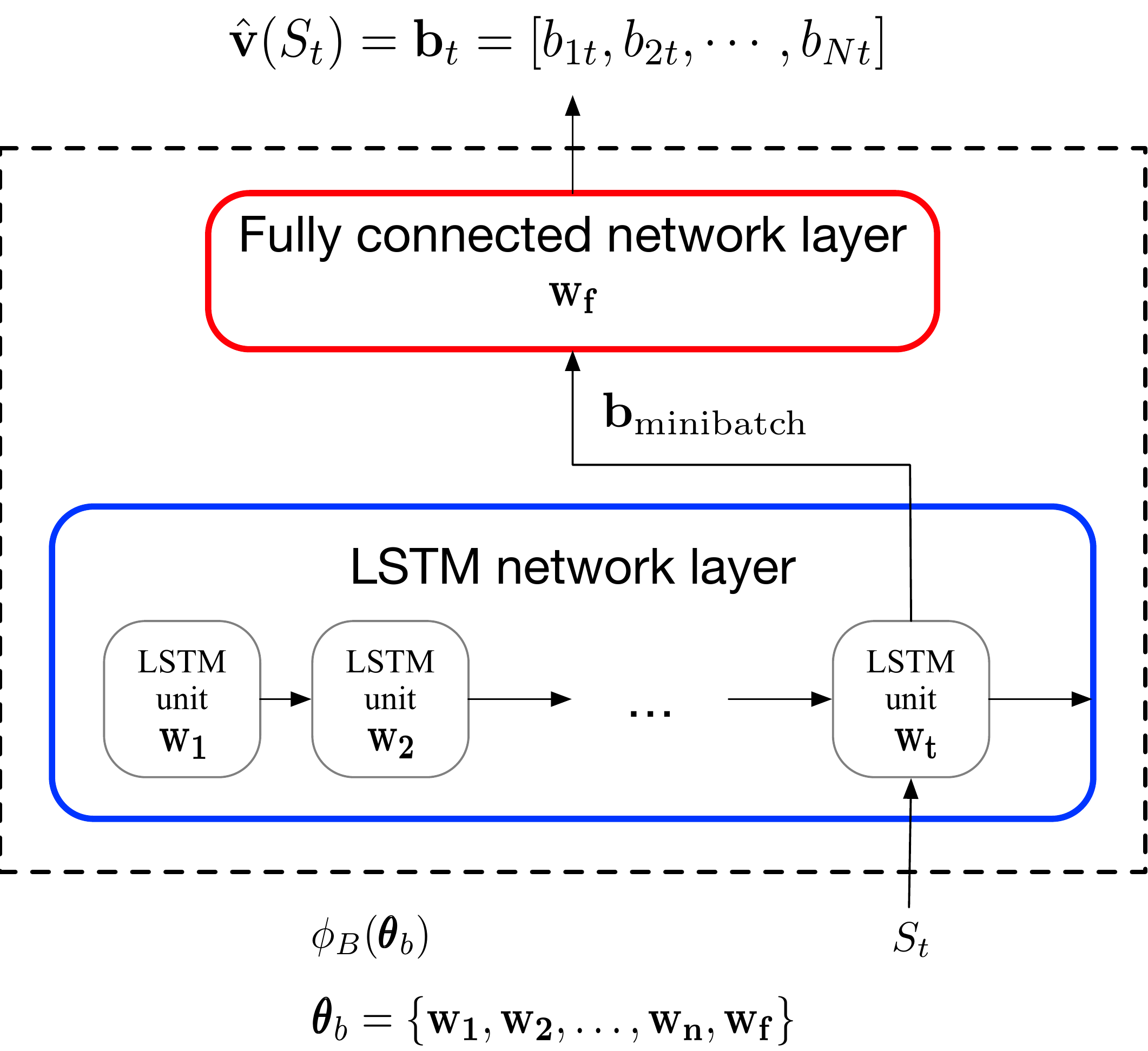}
\caption{LSTM based battery prediction network $\phi_B(\bm \theta_b)$.}
\label{batterylstm}
\end{figure}

In Fig.~\ref{batterylstm}, we illustrate the architecture of our LSTM based
prediction network, which is denoted as the prediction generator.
At the beginning of TS $t$, the BS first observes $S_t$ and imports it
into the LSTM network layer, and the LSTM network outputs the predicted battery
states in multi mini-batch
${\bm b}_{\text{minibatch}}\in \mathbb{R}^{\mathrm{batchsize} \times {N}}$.
Then we use a fully connected network layer to adjust the size of the LSTM output vector
as $\bm b_t \in \mathbb{R}^{1 \times {N}}$.
The BS then announces the access policy based on round-robin.
After UEs' transmissions, the BS receives the true battery
states $\bm R_t$.
We use $\phi_B(\bm \theta_b)$ to denote the prediction generator with the
input $S_t = \{ \bm{X}_t$, $\bm{M}_t,\bm{G}_t \}$ and the output
$\hat {\bm{v}}(S_t) = [ b_{1t},\cdots, b_{Nt} ]$.
Here, $\bm \theta_b$ is the set of network weights, which contains the LSTM
layer parameters $\{\bm w_1, \dots, \bm w_n  \}$ and the fully connected layer parameters
$\bm w_f$, where $n$ is the number of LSTM units.

A good policy in this case is the one minimizing the cumulative discounted
prediction loss based on the observed states.
We utilize experience replay $\mathcal{D}$ (defined in previous Section)
to store the BS experiences
$(S_t, \hat {\bm{v}}(S_t), \bm R_t, S_{t+1})$.
In each TS, we randomly sample a tuple
$(S_j, \hat {\bm{v}}(S_j), \bm R_j, S_{j+1})$
from $\mathcal{D}$ to update the network parameters $\bm \theta_b$.
Stochastic gradient descent is used to minimize the prediction loss by adjusting the
network weight parameter $\bm{\theta}_b$ after each sample
in the direction that would reduce the loss the most.
The weights are updated as
\begin{align}
\label{gradient}
&\bm\theta_b(j+1)  \nonumber \\
&= \bm\theta_b(j) - \frac{1}{2}{\alpha}{\nabla}_{\bm\theta_b(j)}
\left[ \bm V_j - \hat {\bm{v}} \left( S_j, \bm\theta_b(j) \right) \right]^2  \nonumber \\
&=\bm\theta_b(j) + \alpha \left[ \bm V_j - \hat {\bm{v}} \left( S_j, \bm\theta_b(j) \right) \right]
{\nabla}_{\bm\theta_b(j)} {\hat {\bm{v}} \left( S_j, \bm\theta_b(j) \right) },
\end{align}
where $\alpha$ is a positive step size,
$\hat {\bm{v}} \left( S_j, \bm\theta_b(j) \right)$ is
the parameterized prediction values with the network parameters $\bm\theta_b(j)$,
and ${\nabla}_{\bm\theta_b(j)} f(\cdot)$ denotes
the gradient vector of $f$ with respect to $\bm\theta_b(j)$,
and $\bm V_j$ is the target output of the $j$th training step.
We adopt the temporal-difference (TD) policy evaluation
algorithm~\cite{sutton1998reinforcement}, i.e., $TD(0)$,
where the TD error is described as
\begin{align}
\delta_j = \bm R_{j+1} +
\gamma \hat {\bm{v}}_s(S_{j+1}, \bm\theta_b(j)) - \hat {\bm{v}}_s(S_j, \bm\theta_b(j)).
\end{align}
The updating in~\eqref{gradient} can be executed
as $\bm\theta_b(j+1) = \bm\theta_b(j) +
\alpha \delta_j {\nabla}_{\bm\theta_b(j)} {\hat {\bm{v}} (S_j, \bm\theta_b(j))}$.
The battery prediction algorithm based on the deep LSTM network is summarized in
Algorithm 2.
\begin{algorithm}[!t]
	\caption{Deep LSTM based Battery Prediction Algorithm}
	\begin{algorithmic}[1]
		\STATE {Initialize the experience memory $\mathcal{D}$, }
		\STATE {Initialize the parameters of prediction generator network $\phi_B$ with random weights $\bm \theta_b$,}
   \STATE {Initialize the environment and get initial state $S_1$,}
   \STATE {Initialize the total number of episodes $E_p$, }
		\FOR {$t = 1, \cdots, E_p$}
              \STATE {Get prediction output $\hat {\bm{v}} (S_t) = \bm b_t$ given
              by current $\phi_B (\bm \theta_b(t))$,}
              \STATE {Take $ \bm b_t $ as the input of BS access control center,}
              \STATE {BS schedules the UEs according to the round-robin policy,}
              \STATE {BS broadcasts the access policy to all the UEs,
                the selected UEs whose current battery states satisfy
                $P \leq B_{it}$ could complete the transmissions and the others could
                not transmit, }
              \STATE {The BS observes $\bm R_t $
              and the new state $S_{t+1} $, }
              \STATE {Store transition $(S_t,\hat {\bm{v}} (S_t), \bm R_t, S_{t+1})$ in $\mathcal{D}$,}
              \STATE {Sample random mini-batch of transitions $(S_j,\hat {\bm{v}} (S_j), \bm R_j, S_{j+1})$
                      from $\mathcal{D}$,}
              \STATE {Calculate the TD error \\
              $\delta_j = \bm R_{j+1} + \gamma \hat {\bm{v}}_s(S_{j+1}, \bm\theta_b(j)) - \hat {\bm{v}}_s(S_j, \bm\theta_b(j))$,  }
              \STATE {Perform stochastic gradient descent to update network parameters $\bm \theta_b$ based on\\
              $\bm\theta_b(j+1) = \bm\theta_b(j) +
              \alpha \delta_j {\nabla}_{\bm\theta_b(j)} {\hat {\bm{v}} (S_j, \bm\theta_b(j))}$.}
        \ENDFOR
	\end{algorithmic}
\end{algorithm}

\section{Jointly Access Control and Battery Prediction Based on RL}

By jointly considering the access control and battery
prediction, we could relax the requirements of the
knowledge on the UE battery states of the current TS for access control, which
means that only current channel gains are needed at the BS at each TS.

In particular,
we propose a two-layer LSTM based DQN control network, which is applied at
the BS and operates as follows.
At the beginning of each TS $t$, the first LSTM based network layer, which
is used for battery prediction, outputs all the UEs' predicted battery states
based on the history information.
The predicted battery states are then input to the second layer,
which is designed to generate the access control policy.
Next, with the output of the second layer, i.e., the access control policy,
the BS broadcasts the schedule to all the UEs.
Afterwards, the selected UEs execute the policy and transmit their data to the
BS, along with their current true battery states $B_{it}, i \in \mathcal{K}_t$,
which will be stored into the history information for future prediction usage;
those UEs who are not selected remain idle.
The BS finally receives rewards, i.e., the mixture of the sum rate and the
prediction loss.
All the UEs complete the energy conversion and store the energy to the
battery for future use.
The above process repeats in the next TS.

\subsection{Problem Formulation}

The BS operates as the centralized controller that predicts the UE battery
states and schedules the subset of users $\mathcal{K} \in \mathcal{N}$ to
access the uplink channels at each TS.
At TS $t$, the whole system state is denoted as $S_t$.
We deploy a memory component with a window of $W$ at the BS to store the
history data, which contains:
the access scheduling history information, i.e., access indicators,
$\bm{X}_t = \{ I_{ij}\} \in \mathbb{R}^{N \times {W}}$,
where $i \in \mathcal{N}$ and $j \in [t-W+1, t]$;
the history of predicted UE battery information
$\bm{M}_t = \{ b_{ij}\} \in \mathbb{R}^{N \times {W}}$,
where $i \in \mathcal{N}$ and $j \in [t-W+1, t]$;
and the history of the true battery information within the selected UE sets
$\bm{G}_t = \{ B_{ij} \}\in \mathbb{R}^{K \times {W}}$, where
$i \in \mathcal{K}_j$ and $j \in [t-W+1, t]$.
The channel gain at TS $t$ is $\bm{H}_t$.
Thus, we have $S_t = \{\bm{X}_t, \bm{M}_t, \bm{G}_t, \bm{H}_t \}$.
Given $S_t$ and the scheduling policy $\pi$, the action $A_t = \pi(S_t)$
is derived with the DQN control network given in Fig.~\ref{joint}.

Since the performance of the joint solution relies on both the access control
policy and battery prediction results,
the immediate reward contains the received
sum rate $\mathcal{R}_t$ in (7) and the battery prediction
loss $P_{loss}(t)$ in (8),
which is regarded as the penalty to the sum rate.
The immediate reward received at TS $t$ of state $S_t$ by taking
action $A_t$ in the joint solution is set as
\begin{align}
\begin{split}
R_t &= \mathcal{R}_t  - {\beta} P_{loss}(t) \\
&=   \sum_{i \in \mathcal{K}_t} z_{it} F \log \left(1 + \frac{P {H}_{it}}{\sigma^2} \right) -
{\beta} \sqrt{ \sum_{i \in \mathcal{K}_t} {\lvert B_{it} - b_{it} \rvert}^2 },
\end{split}
\end{align}
where $\beta$ denotes the penalty factor for balancing two
different physical quantities.
The long-term system performance, i.e., total discounted reward, from TS $t$
onwards, is given by
\begin{align}
R_t^{\gamma} =& \sum_{k=t}^{\infty}{\gamma^{k-t}}R_{k+1}  \nonumber \\
=& \sum_{k=t}^{\infty}\gamma^{k-t}
\left[ \sum_{i \in \mathcal{K}_{k+1}} z_{i(k+1)} F \log \left(1 + \frac{P {H}_{i(k+1)} }{\sigma^2} \right) \right. \nonumber \\
&\left.-{\beta} \sqrt{ \sum_{i \in \mathcal{K}_{k+1}} {\lvert B_{i(k+1)} - b_{i(k+1)} \rvert}^2 } \right].
\end{align}

The objective of our joint learning algorithm is to obtain the optimal
scheduling policy $\pi$ to maximize the cumulative discounted reward is given by
\begin{align}
\label{eq:ob}
J_3(\pi) = \mathbb{E}(R_t^{\gamma} \mid \pi).
\end{align}
The joint problem can then be formulated as
\begin{subequations}
\label{P3}
\begin{align}
\max\limits_{\pi}\;\;\;
&J_3(\pi)\\
{\rm{s}}{\rm{.t}}{\rm{.}}\;\; & (1)\; \text{and} \;(2) \\
& \sum_{i=1}^N I_{it} = K .
\end{align}
\end{subequations}

\subsection{Two-Layer LSTM-DQN based Joint Network}

\begin{figure}[!t]
\centering
\includegraphics[width=0.9\columnwidth]{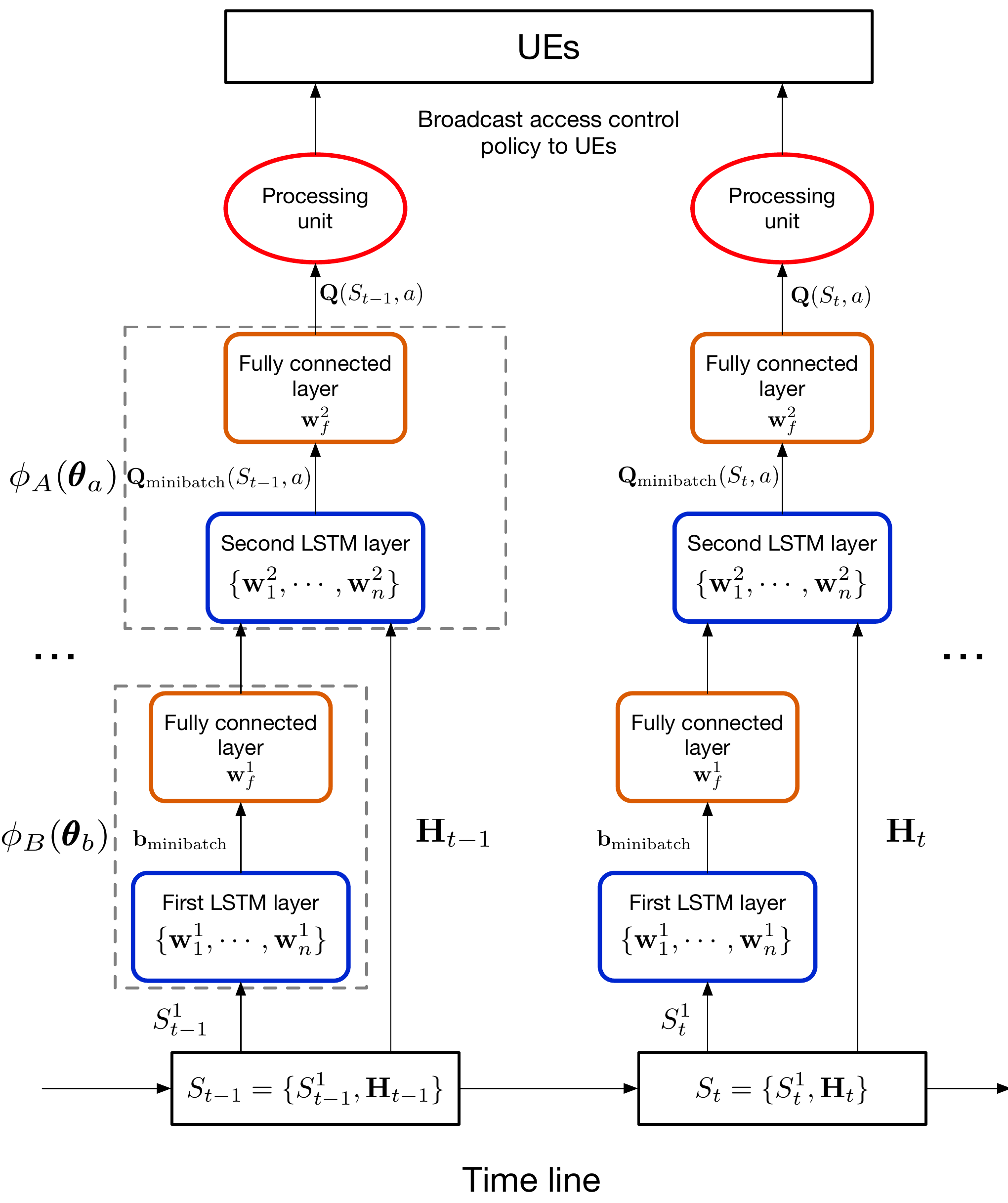}
\caption{Two-Layer LSTM-DQN based control network.}
\label{joint}
\end{figure}

In this subsection, we present the proposed
learning RL network and the algorithm to solve
problem in~\eqref{P3}, where
Fig.~\ref{joint} shows the architecture of the proposed new hybrid control network
combining the LSTM neural network and deep Q-learning enhancement.

The network in Fig.~\ref{joint} can be divided into two layers.
The first is the LSTM layer based network to perform the battery
prediction,
which is called the prediction generator.
In a practical scenario with unknown energy sources, the BS has no information
about the UE EH processes and battery states.
At the beginning of TS $t$, the input of the prediction generator
is the history knowledge within certain time window,
which contains: $\bm{X}_t$, $\bm{M}_t$ and $\bm{G}_t$.
We denote the input as $S^1_{t} = \{ \bm{X}_t$, $\bm{M}_t,\bm{G}_t \}$.
With $S^1_t$, the first LSTM network outputs the predicted battery
states in multi mini-batch
${\bm b}_{\text{minibatch}}\in \mathbb{R}^{\mathrm{batchsize} \times {N}}$.
Then a fully connected network follows to adjust the size of the LSTM
output vector to be the expected $\bm b_t \in \mathbb{R}^{1 \times {N}}$.
We use $\phi_B(\bm \theta_b)$ to denote the prediction generator with the
input $S^1_t$ and the output
$\bm{b}_t = \{ b_{1t},\cdots, b_{Nt}\}$.
Here, the set of network weights $\bm \theta_b$ contains the LSTM
network parameters $\{\bm w^1_1, \dots, \bm w^1_n \}$ and the fully connected
network parameters $\bm w^1_f$, where $n$ is the number of LSTM units.

The second layer is the action generator for producing the access control policy,
which contains an LSTM layer and a fully connected layer.
At TS $t$, the input of the action generator contains: the output values of
$\phi_B$, i.e., $\bm b_t = \phi_B (S^1_t)$; and the current channel states $\bm{H}_t$.
We denote the input of the action generator as $S^2_{t} = \{ \bm b_t, \bm{H}_t\}$.
With $S^2_{t}$, the LSTM layer outputs
the approximated Q-value in mini-batch
$\bm Q_\text{minibatch}(S^2_t,a) \in  \mathbb{R}^{\mathrm{batchsize} \times {L}}$,
where $L$ is the size of the action space with $a \in  \mathcal{A}$.
Then, the fully connected network layer adjusts
the size of the Q-value vector to
$\bm Q(S^2_t, a) \in \mathbb{R}^{1 \times L}$.
Finally, the action generator outputs the approximated
Q-value $\bm Q(S^2_t, a)$.
We represent the action generator as $\phi_A(\bm \theta_a)$ with the
input $S^2_t$ and the output $\bm Q(S^2_t, a) \in \mathbb{R}^{1 \times L}$.
Here, $\bm \theta_a$ is the set of network weights containing the
LSTM layer parameters $\{\bm w^2_1, \cdots, \bm w^2_n\}$ and the fully
connected layer parameters $\bm w^2_f$.

Therefore, by combining the prediction generator and action generator, the
proposed two-layer LSTM-DQN based joint network can be represented as
$\phi_A ( \phi_B ( \bm{X}_t, \bm{M}_t,\bm{G}_t ), \bm{H}_t )$ with the
entire input as $S_t = \{\bm{X}_t, \bm{M}_t, \bm{G}_t, \bm{H}_t \}$
and the output approximation of Q-value as $\bm Q(S_t, a) = \bm Q \left(\{\phi_B(
\bm{X}_t, \bm{M}_t,\bm{G}_t), \bm{H}_t \}, a \right)$.

In the proposed RL network, we learn the parameters $\bm \theta_b$ of the prediction
generator and $\bm \theta_a$ of the action generator jointly.
The parameters of the two-layer joint network is denoted
by $\bm \theta = \{ \bm \theta_a, \bm \theta_b \}$.
At the beginning of TS $t$, the BS receives $S_t$.
The prediction generator $\phi_B$ firstly outputs $\bm b_t$,
and the BS then stores $\bm b_t$ by updating its history memory.
With the predicted battery states and channel gains, the action generator then
outputs the Q-value $\bm Q(S_t, a)$.
As explained in Section III, the balance between the exploration of new actions
and the exploitation of the known actions is realized by the $\epsilon$-greedy
action selection method~\cite{narasimhan2015language}.
With $\epsilon$-greedy, the BS either takes actions randomly
with probability $\epsilon$ or follows the greedy policy (chosing the action $A_t$
by $\max_{A_t \in \mathcal{A}} \bm Q(S_t, A_t)$) with probability $1-\epsilon$,
where $0 < \epsilon <1$.

After executing the selected action $A_t$, the BS receives the immediate
reward $R_t$.
We keep tracking the BS's previous experience in a replay memory data set
$\mathcal{D} = \{e_1, \cdots , e_t\}$, with
$e_t = (S_t, A_t, R_t, S_{t+1})$.
Instead of performing updates to the Q-values using transitions from the
current episode, we sample a random transition
$(\tilde s, \tilde a, \tilde r, \hat s)$ 
from $\mathcal{D}$~\cite{mnih2015human}.
Following the Q-learning approach, we obtain the target Q-value
$y_t = \tilde r + \gamma \max_{\hat a} Q(\hat s, \hat a; \bm \theta^-)$,
where $Q(\hat s, \hat a; \bm \theta^-)$ is the parameterized approximate
value function with network parameters
$\bm \theta^{-} =  \{ \bm \theta_a^{-}, \bm \theta_b^{-} \} $ obtained
from the previous iteration.

We can get the following loss function $\mathcal{L}_t(\bm \theta)$ to minimize:
\begin{align}
\mathcal{L}_t(\bm \theta) =& (y_t - Q(\tilde s, \tilde a ; \bm \theta))^2  \nonumber \\
=& \left( \tilde r + \gamma \max_{\hat a} Q(\hat s, \hat a; \bm \theta^-) -
Q(\tilde s, \tilde a ; \bm \theta)  \right)^2.
\end{align}
The updates on $\bm \theta$ can be performed using
the stochastic gradient
of ${\nabla}_{\bm \theta}{\mathcal{L}_t(\bm \theta)}$.
It is worth mentioning that, in this gradient, $\bm \theta$ contains both
two-layer network parameters, i.e., $\bm \theta = [\bm \theta_{a}, \bm \theta_{b}]$,
which means that the prediction generator and access control policy network are
trained in a joint way~\cite{cho2014learning}\cite{mikolov2011extensions}.
The overall joint access control and battery prediction algorithm is
summarized in Algorithm~3.

\begin{algorithm}
\caption{Algorithm for Joint Problem~\eqref{P3}}
  \label{table:2}
\begin{algorithmic}[1]
	\STATE {Initialize the experience memory $\mathcal{D}$, }
	\STATE {Initialize the parameters of prediction generator network $\phi_B$ with random weights $\bm \theta_b$,}
	\STATE {Initialize the parameters of action generator network $\phi_A$ with random weights $\bm \theta_a$,}
	\STATE {Initialize the total number of episodes $E_p$,}
	\FOR{$eposode = 1, \cdots, E_p$}
          \STATE {Initialize the environment and get initial observation state $S_1 =
          \{ S^1_t, \bm{H}_1 \}$, $S^1_t = \{ \bm{X}_1,\bm{M}_1,\bm{G}_1 \} $, }
          \FOR{$t=1, \cdots, T$}
             \STATE {Get the predicted battery states $\bm b_t = \phi_B (S^1_t)$,}
             \STATE {Get the input state $S^2_t = \{\phi_B (S^1_t), \bm{H}_t \}$ for $\phi_A$, }
             \IF{$random () \leq \epsilon$}
               \STATE {Select a random action $A_t \in \mathcal{A}$;}
             \ELSE
               \STATE {Compute $Q(S_t, a)$ for all actions using $\phi_A$,}
               \STATE {Select $A_t = \underset{{a \in \mathcal{A}}}{\arg\max} Q(S_t, a; \bm \theta)$.}
             \ENDIF
             \STATE {Execute $ A_t $, observe reward $ R_t $,}
             \STATE {Get new state $ S_{t+1} = \{ S^1_{t+1}, \bm{H}_{t+1} \}$, }
             \STATE {Store transition $(S_t, A_t, R_t, S_{t+1})$ in $\mathcal{D}$,}
             \STATE {Sample random mini-batch of transitions
             $ (\tilde s, \tilde a, \tilde r, \hat s) $  from $\mathcal{D}$,}
             \STATE {Set $y_t = \tilde r$ if $t+1$ is the terminal step of the
             episode ($t+1=E_p$);
              otherwise, $y_t = \tilde r + \gamma \max_{\hat a} Q(\hat s, \hat a; \bm \theta_a^-)$, }
             \STATE {Perform the stochastic gradient descent step on the loss function
             $\mathcal{L}_t(\bm \theta_a) = (y_t - Q(\tilde s, \tilde a ; \bm \theta_a))^2$
             to update the network parameters $\bm \theta$.}
           \ENDFOR
      \ENDFOR
\end{algorithmic}
\end{algorithm}

\section{Simulation Results}
In this section, we demonstrate the performance of the two proposed RL based algorithms
by simulations.
All the results are performed in a simulated LTE uplink scenario with one BS
and 30 randomly walking UEs with the speed of $1 \text{m/s}$.
The UEs' energy arrival processes are modeled as Possion arrival processes
with different arrival rates.
The cell range is $500\text{m} \times 500\text{m}$ with
$\text{pathloss} = 128.1+37.6 \log(d)$, where $d$ is the transmission distance
~\cite{zhang2017connecting}.
The system total bandwidth is $F = 5MHz$ and the penalty factor $\beta$ is $10^2$.
All the battery states are quantized into integer units with 5dBm per unit and
the transmit power is at 2 units.

The LSTM network consists of 128 units and the fully connected layer uses a
tanh activation function.
The learning rate $\alpha$ is fixed as $ 10^{-4} $
and the discount factor $ \gamma $ is set to be $ 0.99 $.
We train the deep RL network with a mini-batch size of 16 and a replay
buffer size of $ 10^5$.
All simulation results are obtained based on the deep
learning framework in TensorFlow 1.2.1.

To compare the performance of our proposed algorithms, we consider the
following alternative approaches:
1) an offline benchmark provides an upper bound where the BS is assumed to have
perfect non-causal knowledge on all the random processes;
2) a myopic policy (MP), which is a widely used data-driven approach in the
multi-armed bandit model~\cite{blasco2015multi};
3) the round-robin scheduling; and 4) random scheduling.
It is worth mentioning that the presented rewards are averaged by taking the mean
over a fixed moving reward subset with a window of 200 training steps
to achieve smoother and more general performance
comparison.

\begin{figure}[!t]
\centering
\includegraphics[width=0.9\columnwidth]{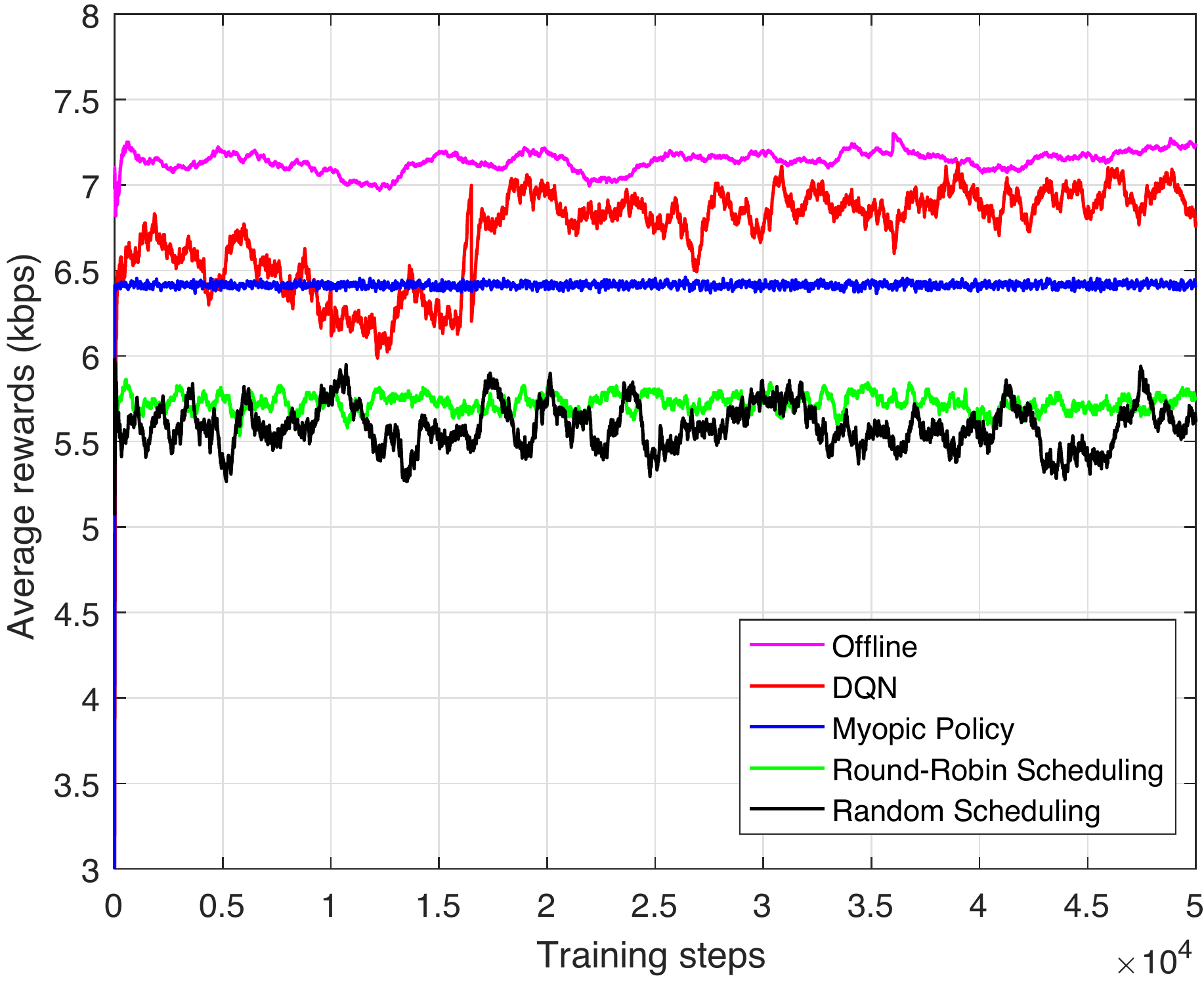}
\caption{Average reward vs. training steps.}
\label{fig1}
\end{figure}

\begin{figure}[!t]
\centering
\includegraphics[width=0.9\columnwidth]{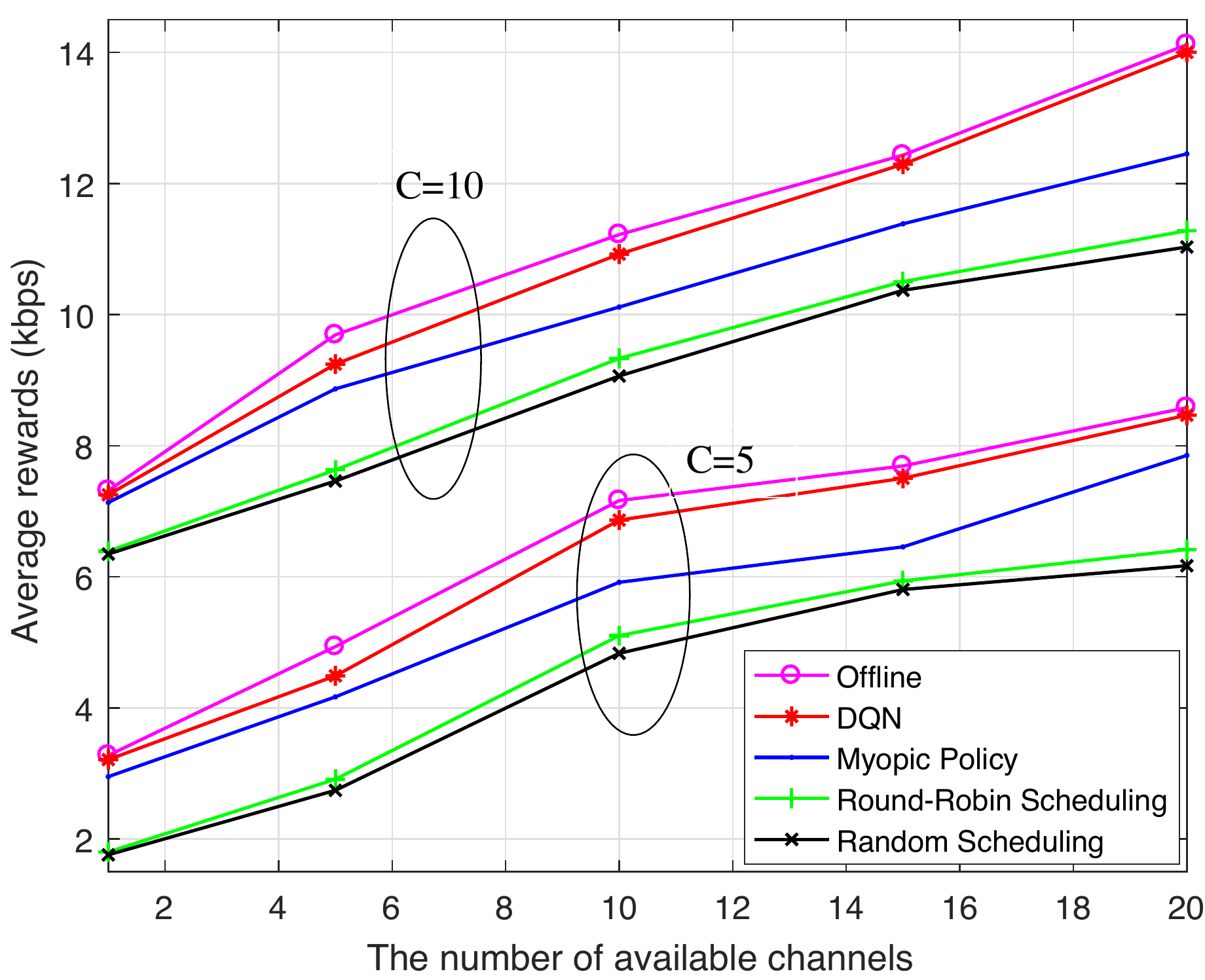}
\caption{Average reward vs. battery capacity and available channels.}
\label{fig2}
\end{figure}

Firstly, we investigate the performance of Algorithm~1 compared with other methods.
As shown in Fig.~\ref{fig1}, the proposed learning algorithm always achieves
a higher average reward than round-robin and random scheduling.
This is intuitive since round-robin only considers the access fairness among
all the UEs, and random selection makes the decision even more blindly.
With the increase of training steps, the proposed DQN scheme at first stays in an
unstable exploration stage.
Then, it gradually outperforms MP after $1.8 \times 10^4$
training steps.
Finally, it converges and becomes stable.
This is because that MP always focus on the current sum rate optimization
based on the battery beliefs\cite{blasco2015multi}, while the DQN algorithm takes
the long-term performance into consideration, resulting in a more efficient
resource utilization and higher sum rate.
Furthermore, we observe that the performance gap between our proposed algorithm
and the offline upper bound gets smaller as the training step increases.
We also compare the average sum rates under different numbers of
available channels and battery capacities.
We can also see from Fig. 6 that after getting
stable, the average reward of the proposed DQN based algorithm
achieves a value that is $11.93\%$ higher than that with the MP approach,
$19.47\%$ higher than the round-robin
approach, and $26.05\%$ higher than the random scheduling approach.
Meanwhile, the average reward of the proposed DQN based approach is
$3.3\%$ lower than the upper bound (i.e., the offline scheduling).
It can be seen in Fig.~\ref{fig2} that the proposed DQN algorithm always beats the
counterparts
when the number of available channels changes from 2 to
20 and the battery capacity changes from 5 to 10 units.
Furthermore, the average sum rate increases with the number of available channels.
By increasing the battery capacity from 5 to 10 units, the battery
overflow is reduced, which results in a higher average sum rate.

\begin{figure}[!t]
\centering
\includegraphics[width=0.9\columnwidth]{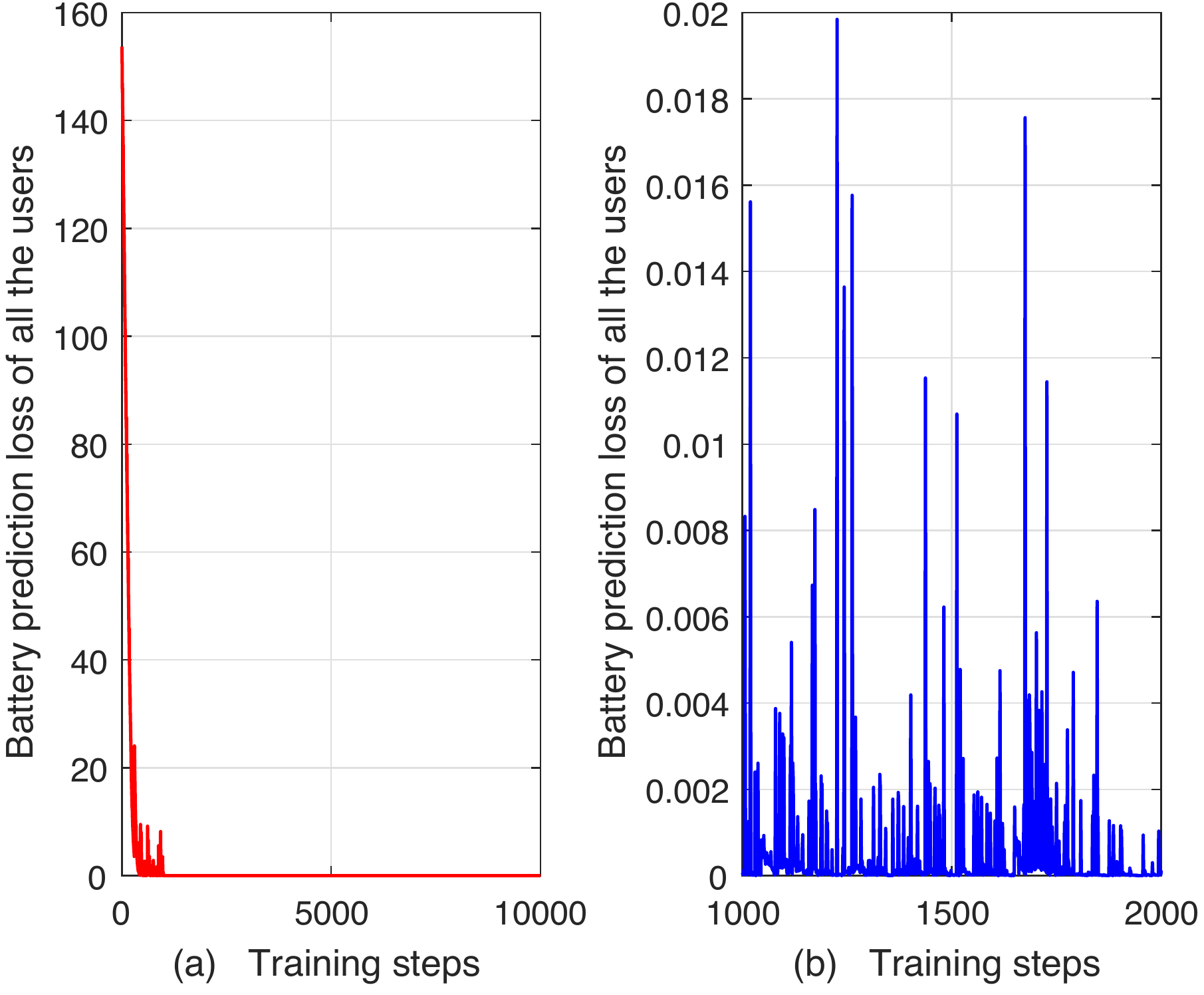}
\caption{Battery prediction loss vs. training steps.}
\label{fig3}
\end{figure}

\begin{figure}[!t]
\centering
\includegraphics[width=0.9\columnwidth]{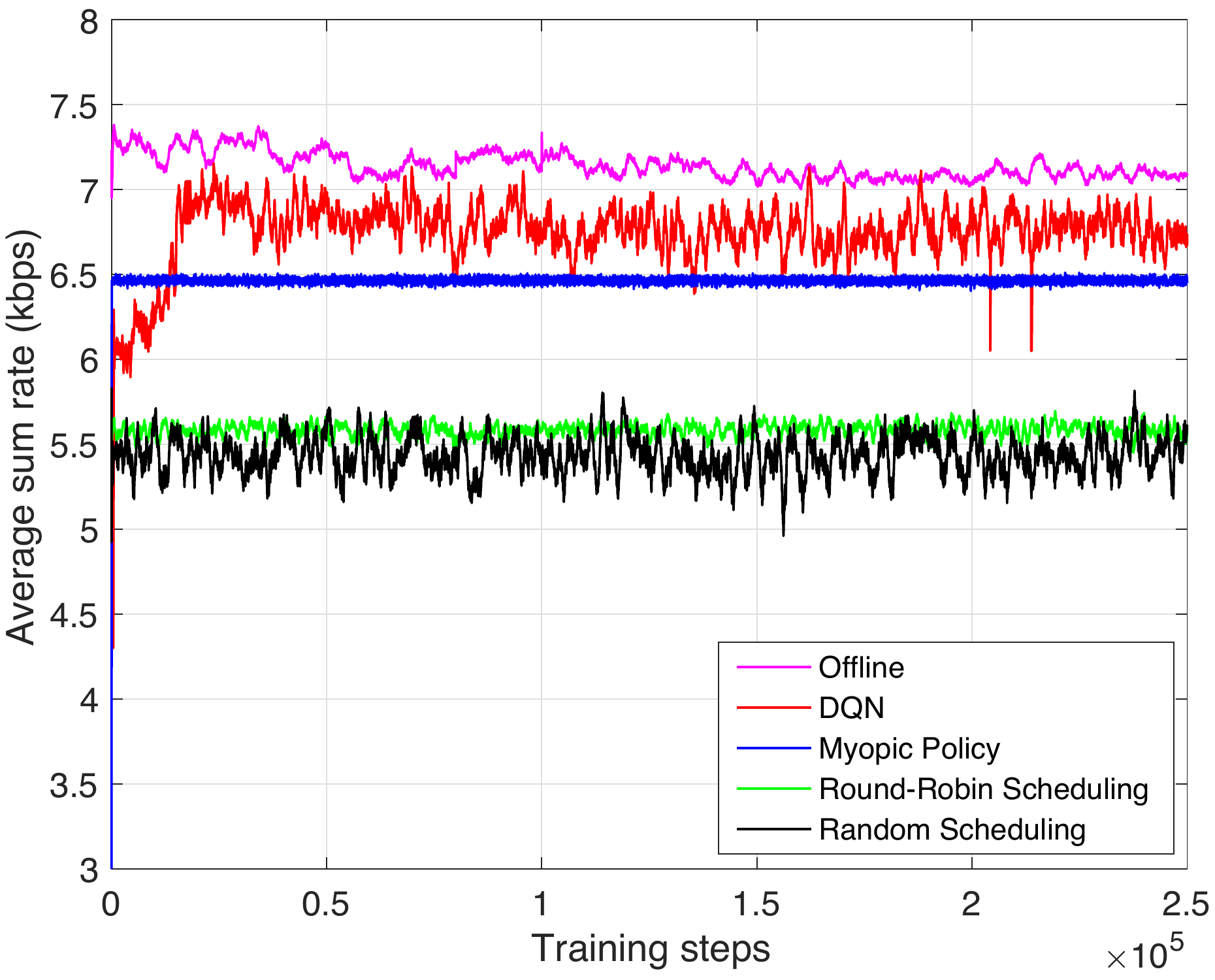}
\caption{Average sum rate vs. training steps.}
\label{fig4}
\end{figure}

For Algorithm~2, we perform simulations for the total battery prediction
loss of all the UEs with the proposed DQN algorithm, shown in Fig.~\ref{fig3}.
It can been seen from Fig.~\ref{fig3}(a) that the prediction
loss is quite large at the beginning.
With the increase of training steps, the loss becomes
smaller and goes to a stable value after about 1000 training steps.
We zoom in over the loss values between 1000 and 2000 training steps in Fig.~\ref{fig3}(b).
The average UE battery prediction loss shown in
Fig.~\ref{fig3} is about 0.0013 units.
It is obvious that the prediction loss is small enough and the proposed deep LSTM
prediction network provides good prediction performance.

\begin{figure}[!t]
\centering
\includegraphics[width=0.9\columnwidth]{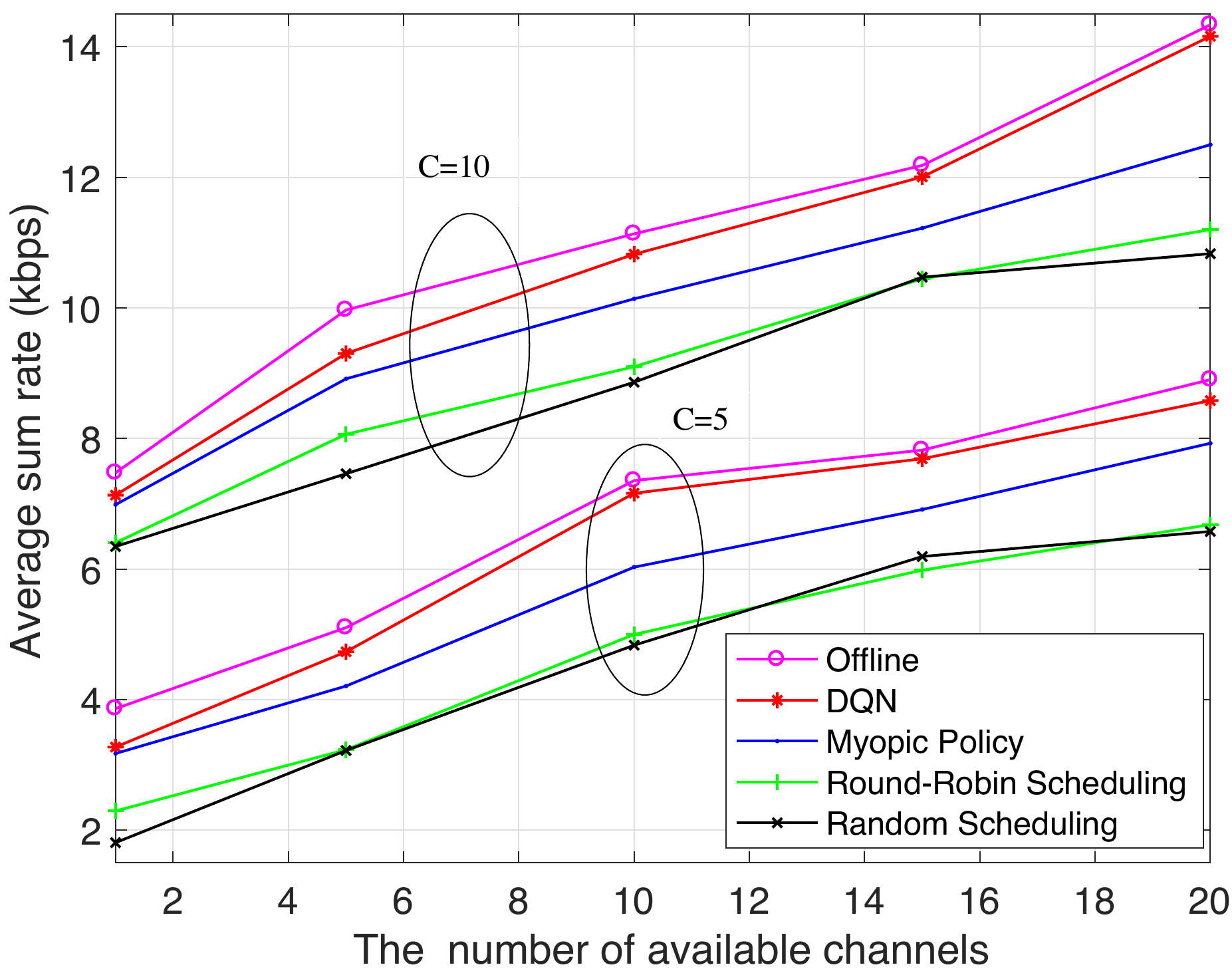}
\caption{Average sum rate vs. battery capacity and available channels.}
\label{fig5}
\end{figure}

\begin{figure}[!t]
\centering
\includegraphics[width=0.9\columnwidth]{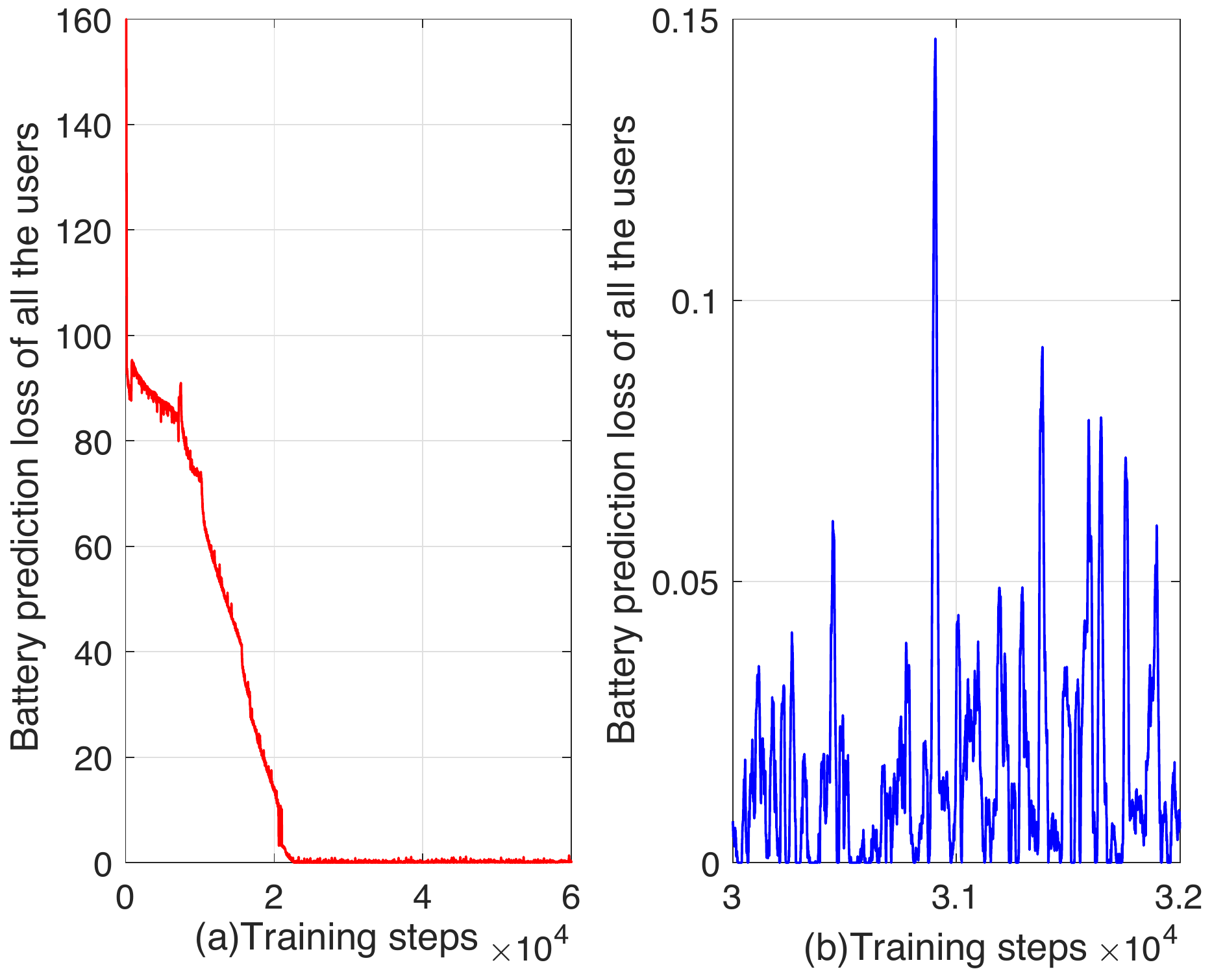}
\caption{Battery prediction loss vs. training steps.}
\label{fig6}
\end{figure}

At last, the performance of the proposed joint scheme in Algorithm~3
is investigated.
The average sum rate
and the corresponding battery
prediction loss are shown in Fig.~\ref{fig4}~-~Fig.~\ref{fig6}, respectively.
It can be seen from Fig.~\ref{fig4} that the data-drive approaches, i.e., MP and
DQN, always perform better than the round robin and random scheduling, which
is intuitive and obvious since the last two have no consideration over sum rate
optimization.
The proposed DQN based algorithm stays in an exploration stage at the beginning,
which is unstable and resulting in a worse performance compared with
the MP approach.
With more training steps, as expected, the average sum rate
of the proposed
DQN algorithm arises to be better than MP and remains stable after about
$2.2 \times 10^4$ training steps.
Compared with the offline upper bound, we can observe that
though the average sum rate of the proposed DQN cannot achieve the upper bound, the performance
gap between the two gets smaller as the training step increases.
It can be seen in Fig. 9 that after getting stable,
the average reward of the proposed joint algorithm
is $9.96\%$ higher than that of the MP approach, $21.1\%$ higher than the round-robin
approach, $24.7\%$ higher than the random scheduling approach, and
$4.14\%$ lower than the offline scheduling.
The average sum rates under different numbers of
available channels and battery capacities are shown in Fig.~\ref{fig5}.
It can be observed that the proposed DQN algorithm always defeats the MP, round-robin
and random approaches when the number of available channels changes from 2 to
20 and the battery capacity changes from 5 to 10 units.
Besides, it is obvious that the average sum rate of the proposed DQN algorithm is
close to the upper bound.
Furthermore, we see that the average sum rate increases with the increase
of battery capacity and the number of available channels, owing to the
reduction of battery overflow.

The performance of the corresponding battery prediction network for the joint
scheme is shown in Fig.~\ref{fig6}.
Compared with Fig.~\ref{fig4}, we see that the battery prediction loss
goes to a stable stage earlier than the average sum rate.
This is because that the output of battery prediction network is the main
input for the access control network;
only after the battery prediction is accurate enough,
the BS could generate good scheduling policies that achieve high sum rates.
It can been seen from Fig.~\ref{fig6} (a) that the prediction
loss is quite large at the beginning, becomes
smaller as the training step increases, and gets stable after about 22000
training steps.
We zoom into the loss values from the 
30000th to 32000th
training steps
in Fig.~\ref{fig6} (b).
The average UE battery prediction loss shown in
Fig.~\ref{fig6} is about 0.0175 units.
It is obvious that the prediction loss is small enough and the proposed deep LSTM
prediction network provides good prediction values.

\section{Conclusion}

In this paper, we developed three RL based methods to solve the user
access control and battery prediction problems in a multi-user EH based
communication system.
With only causal information regarding the channel
and UE battery states, the LSTM-DQN based scheduling algorithm was designed to find
the optimal policy with the objective of maximizing the long-term discounted
uplink sum rate, driven by only instantaneous system information.
The battery state prediction algorithm based on deep LSTM was proposed to
minimize the prediction loss.
Furthermore, the joint problem was considered,
where we proposed a two-layer LSTM based network which is trained jointly
with deep Q-learning to maximize the long-term discounted sum rate
and minimize
the cumulative battery prediction loss simultaneously.
The simulation results under different conditions were also provided to
illustrate the effectiveness of our proposed RL based methods.


\bibliographystyle{IEEEtran}
\bibliography{IEEEabrv,refs.bib}

\end{document}